\documentclass{aa}

\usepackage{graphicx}

\begin{document}

   \title{A catalogue of dust clouds in the Galaxy}

   \author{C.M. Dutra\inst{1,2}, E. Bica\inst{2}}

   \offprints{C.M. Dutra -- dutra@iagusp.usp.br}

 \institute{Instituto Astron\^omico e Geof\'\i sico da USP, CP\, 3386, S\~ao Paulo 01060-970, SP, Brazil\\
\mail{}
\and
Universidade Federal do Rio Grande do Sul, IF, CP\,15051, Porto Alegre 91501--970, RS, Brazil\\
\mail{}
}

\date{Received ; accepted }

\abstract{
   In this study 21 catalogues of dust clouds in the Galaxy were cross-identified by taking into
account  available properties such as position, angular dimensions, opacity class and velocity.  An
initial list of $\approx$6500 entries was condensed into a cross-identified all-sky catalogue  
containing 5004 dust clouds.  
In particular, the  transition zone between high and low Galactic
latitude studies was also cross-identified.  The unified  catalogue  contains 525 high-latitude clouds. 
The catalogue deals primarily with optical dark nebulae and globules, but it includes as well substantial information from their molecular counterparts. Some previously uncatalogued  clouds were detected
on optical images and FIR maps. Finally, we address recent results and prospective work 
based on NIR imaging,  especially for  clouds  detected in  the 2MASS $K_s$ Atlas.    
}

\titlerunning{Dust clouds in the Galaxy}

\authorrunning{Dutra \& Bica} 

\maketitle

%


\section{Introduction}

Dust clouds are fundamental to understand many  issues in the Galaxy.  
Optical obscuration   is an obstacle to the study of Galactic structure from the perspective of
star clusters and H\,II Regions.
However, dust association to molecular gas, and velocity determinations  (e.g. Blitz et al. \cite{bli}, Dame et al. \cite{dam}) make them  excellent tracers of related structures in the Galaxy. Since some clouds  are actively forming stars, with protostellar objects and/or  embedded  clusters (e.g.  Lawson et al. \cite{law}, Lada et al. \cite{lad2}), clouds are  interesting targets to search for  new IR clusters. Near infra-red (NIR) surveys such as the Two Micron All Sky Survey (2MASS -- Skrutskie et al. \cite{skr}) are opening this possibility (e.g. Dutra \& Bica \cite{dut}). 

 Much information on dust clouds remains strewn throughout many catalogues, and subsequent studies 
often refer to a particular designation. A unified catalogue would be useful for searches of new clouds, especially for surveys in  wavelengths other than optical. Similar to catalogue 
efforts in other fields, such as those dealing with  galaxies, open clusters or globular clusters,
the present study aims at providing a step forward to  an overall  catalogue of dust clouds. In Sect. 2 
input catalogues  are gathered. In Sect. 3 merging and cross-identification
procedures are given, and results are presented.  
In Sect. 4  some catalogue properties  and prospective work are discussed.

\section{Input catalogues}

Input catalogues dealing mostly with low-latitude clouds ($|b|$ $<$ 25$^{\circ}$), their acronyms  and  number of entries (prior to cross-identifications) are given in Table 1. Barnard (\cite{bar}, \cite{bar2}) first catalogued  clouds in the northern and equatorial zones. Lynds (\cite{lyn}) built  the largest set  of dark clouds (LDN) north of $\delta$ = -33$^{\circ}$  using the Palomar Sky Survey (PSS), including  some high-latitude clouds. Sandqvist and Lindroos (\cite{san}) detected  clouds  in the PSS for -33$^{\circ} < \delta <$ -46$^{\circ}$. Sandqvist (\cite{san2}) found   clouds  south of $\delta$ = -42.5$^{\circ}$ using the ESO (B) Atlas. Bernes (\cite{ber}) looked for  bright nebulae mainly related to  LDN's, and reported some new dust clouds. Zealey et al. (\cite{zea}) compiled  cometary globules -- as a rule 'bright dark' nebulae with reflection and absorption components. 
Feitzinger \& St\"uwe (\cite{fei}) inspected  ESO/SERC plates and provided two catalogues: 
(i) extended  clouds, and (ii) globules ($\approx$ 6$^{\prime}$ or less). Hartley et al. (\cite{har}) presented the largest southern set of clouds from a search on ESO/SERC J plates for $\delta$ $\leq$ -35 $^{\circ}$. 

Some Table 1 catalogues  deal with dense cloud cores and isolated small clouds. 
Myers et al. (\cite{mye}) identified  small clouds using  the PSS  for  low-mass star formation studies by means of CO.  Clemens \& Barvainis (\cite{cle}) used  the PSS for small cloud searches  and studied their optical, IR and millimeter properties. Vilas-Boas et al. (\cite{vil}) investigated small clouds  from  ESO/SERC J plates, extinction maps and previous studies, located in  nearby  clouds or complexes such as Coalsack and Vela. Bourke et al. (\cite{bou}) analysed  clouds with opacity class A from Hartley et al., two of them new. Dense clouds can  harbour embedded cores of Young Stellar Objects (YSO) or Pre-Main-Sequence (PMS) stars. Parker (\cite{par}) provided accurate positions for 147 Lynds$^{\prime}$ clouds  with opacity class 6. Lee \& Myers (\cite{lee})  listed the largest set of  dense cores, with  opacity classes 5 and 6 from Lynds and A  from Hartley et al. Parker (\cite{par}) and Lee \& Myers (\cite{lee}) studied  the relation of dense cores  to  IRAS sources, and in turn to  YSO and PMS. Finally, Vilas-Boas et al. (\cite{vil2}) studied cores in Lupus, Corona Australis, Scorpius  and Vela.

 The input data were complemented with objects from individual studies or small lists: 
(i) 21 nearby clouds and complexes (Cambr\'esy \cite{cam}) with  extinction maps from star counts;
(ii) 23 nearby  clouds and complexes as studied in CO (Dame et al. \cite{dam}); (iii)  
molecular clouds from individual studies such as TMC-1 and TMC-2 (Churchwell et al. \cite{chu}), 
OMC-1 (Ziurys et al. \cite{ziu}), OMC-2 and 
OMC-3 (Chini et al. \cite{chi}), Heiles 2 (Heiles \cite{hei}) and  
MT 1 (Maddalena \& Thaddeus \cite{mad});  (iv) 2
cometary globules -- CG --  and 7 Gum Nebula Dark Clouds --GDC  (Reipurth \cite{rei}); and 
finally, (v) a
pre-cometary globule (Lefloch \& Lazareff \cite{lef}). The total number of 
entries from catalogues and lists preferentially including  low-latitude clouds is 5654.

\begin{table}
\caption[]{Input catalogues: mostly low latitudes}
\begin{scriptsize}
\renewcommand{\tabcolsep}{0.9mm}
\begin{tabular}{lcc}
\hline\hline
Catalogue&Acronym&Entries\\ 
\hline 
Barnard (1919, 1927)&B&349\\
Lynds (1962)&LDN&1806\\
Sandqvist \& Lindroos (1976) & SLDN&42\\
Sandqvist (1977)&SDN&95\\
Bernes (1977)&BDN&81\\
Zealey et al. (1983)&CG&34\\
Myers, Linke \& Benson (1983)&MLB&90\\
Feitzinger \& St\"uwe (1984)$^{\rm a}$&FeSt1-&489\\
Feitzinger \& St\"uwe (1984)$^{\rm b}$&FeSt2-&331\\
Hartley et al.(1986)&HMSTG&1101\\
Clemens \& Barvainis(1988)&CB&248\\
Parker (1988)&P&147\\
Vilas-Boas et al. (1994)&VMF$^{\rm c}$&101\\
Bourke, Hyland \& Robinson (1995)&BHR&169\\
Lee \& Myers (1999)&LM&406\\
Vilas-Boas et al. (2000)&$^{\rm d}$&104\\
\hline
\end{tabular}
\end{scriptsize}
\begin{list}{}
\item  Notes: $^{\rm a}$ extended clouds; $^{\rm b}$ globules; 
$^{\rm c}$ additional designation is the 
related  dust complex abbreviation plus running 
number; $^{\rm d}$ designation as in $^{\rm c}$. 
\end{list}
\end{table}

  The acronyms and number of entries of the input 
catalogues dealing mostly with  high-latitude clouds are given in Table 2. Most of them deal with  CO studies, but searches based on far infra-red (FIR) dust emission atlases such as IRAS provided most entries. High-latitude 
clouds present low visual extinction (translucent -- van Dishoeck et al. \cite {van}) and are therefore difficult to be detected on  photographic surveys.  
Magnani et al. (\cite{mag}) and Keto \& Myers (\cite{ket}) studied high-latitude clouds in CO (the latter 
study includes some lower-latitude clouds). Magnani et al. (\cite{mag2}) 
compiled list of clouds taken from the literature, providing properties and references
for clouds with the acronyms  UT , ir, HSVMT, G, Stark and 3C. The 
acronym HRK stands for clouds in  regions studied in  HI, CO and IR (Heiles et al. \cite{hei3}).
FIR emission excess with respect to H\,I  indicates cold clouds where hydrogen appears in molecular form 
(de Vries et al. \cite{vri}).  D\'esert et al. (\cite{des}) used the IRAS 100 $\mu$m data coupled to the  Berkeley H\,I survey (Heiles \& Habing \cite{hei2}) to detect 516 infrared excess clouds (IREC) for  $|b|$ $>$ 5$^{\circ}$. Reach et al. (\cite{rea}) used  DIRBE/COBE  in conjunction with the Leiden-Dwingeloo H\,I survey (Hartmann \& Burton \cite{han}) to create higher resolution  maps. From the IR excesses they retrieved 60 previous clouds and found 81 new ones (DIR -- Diffuse Infrared Clouds). Also included were 2
 new  clouds studied in CO by  Hartmann et al.
(\cite{vri}), and the  absorption component of the Draco Nebula (Goerigk et al. \cite{goe}).

\begin{table}
\caption[]{Input catalogues: mostly high latitudes}
\begin{scriptsize}
\renewcommand{\tabcolsep}{0.9mm}
\begin{tabular}{lcc}
\hline\hline
Catalogue&Acronym&Entries\\ 
\hline 
Magnani et al. (1985)~~~~~~~~~~~~~~~~~~~&MBM&57\\
Keto \& Myers (1986) & KM& 18\\ 
D\'esert et al. (1988)&IREC&~516\\
Magnani et al. (1996)&various& 120\\ 
Reach et al. (1998)&DIR&141\\ 
\hline
\end{tabular}
\end{scriptsize}
\end{table}
\section{Catalogue construction and contents} 

The procedures  for catalogue construction and cross-identifications follow those outlined 
for the LMC and SMC  extended objects (Bica \& Schmitt \cite{bic}, Bica et al. \cite{bic1}). 
Available electronic catalogues were retrieved from CDS and 
remaining ones were typed. The original coordinates (usually either  B1950.0 $\alpha,\delta$ 
or in $\it l,b$) were transformed to J2000.0 $\alpha,\delta$ and to homogeneous $\it l,b$ values. 
All catalogues were merged into one file and sorted  by $\it l$. Equivalent objects 
were merged into a single  line considering  positions and angular dimensions.
This was complemented with available properties such as optical opacity class. Previous 
cross-identifications were examined. Revisions occurred and many  new equivalences
were found. Doubts on optical clouds were inspected on  Digitized Sky Survey (DSS and XDSS) 
images and/or sky charts generated by means of  the Guide Star Catalogue. Recent catalogues dealing with small nebulae (Sect. 2) and that by Hartley et al. (\cite{har}) produced accurate coordinates, which is not always the case of  early works. 

   Studies in Sect. 2 and Hilton \& Lahulla$^{\prime}$s (\cite{hil})   
provided  distances from several methods 
(e.g. related stars or OB complexes, background and foreground stars, kinematical). Studies in Sect.2 
and Otrupcek et al.$^{\prime}$s (\cite{otr})  provided LSR CO (in a few cases H\,I) velocities. 
Velocity and distance estimates  were  additional contraints for the definition of clouds and their physically related groups.
      
   Opacity classes were reduced to Lynds' (\cite{lyn}) scale. Internal 
variations in some clouds were averaged. For detailed 
values we refer to the original studies. Optical opacity class has been  
a useful parameter for cloud classifications and selections.
We  provide FIR contrast parameters  for similar purposes.
Schlegel et al. (\cite{sch})  built an all-sky reddening $(E(B-V))_{FIR}$ map based   
on 100 $\mu$m dust thermal emission with resolution of $\approx$ 6$^{\prime}$, and temperature corrections 
based on lower resolution DIRBE maps.
We extracted  the  reddening central value   in the direction
of each  cloud and  4 background surrounding positions  distant
1.3$\times$  the cloud's major axis dimension for those larger than  10$^{\prime}$.
For smaller clouds the background distance was 2$\times$ the major axis dimension.  We 
computed  the background  average value, its  fluctuation, and  2 contrast parameters. 
$\Delta$ is defined as the difference  between centre and  
background values, and  $\rho$ as the ratio of these quantities. 
Note that $\Delta$ can present negative  
and $\rho$ $<$ 1 values,  since background
contributions  can be substantial, especially as one approaches
the Plane (Dutra \& Bica \cite{dut2}). Neighbouring clouds, position uncertainties, irregular 
shapes and  extensions can also disturb. Although  clouds
are concentrated to the Plane $\approx$ 3750 clouds  present  positive FIR contrast
($\Delta$ $> 0$  and  $\rho$ $>$ 1). Selections
for a variety of studies can be obtained from such parameters, from candidates
to line-of-sight near projections to isolated dense clouds. As cautionary remark,
the present discrete cloud FIR values and parameters should not be  taken as 
true dust column densities and their combinations,  especially 
for small clouds owing to the low resolution of the temperature maps,
coupled to other limitations.

Some newly identified clouds are included: (i)
Magnani et al. (\cite{mag3})  detected CO  along 133 directions for $b$ $<$ -30$^{\circ}$, 
58 were new detections and 75 were related to  26 previous clouds. 
We compared Reach et al.'s (\cite{rea}) FIR maps with  CO detections by  
Magnani et al. (\cite{mag3}). As a result we found evidence of  16 new clouds
indicated by the acronym CODIR (one additional cloud turned out to be IREC 267); (ii)
We were not particularly searching for new clouds in the present study, but during the 
cross-identification process and inspections of optical images and FIR maps  
we found 15 uncatalogued clouds (DBDN). They are 
2 low-latitude globules, 10  high-latitude and 3  transition zone extended clouds.
(iii) The  dust clouds recently reported by Dutra \& Bica (\cite{dut4}) using
the 2MASS Atlas. These 5 clouds (DBIRDN) are opaque in the $K_s$ band
and are projected onto the central bulge. Some of them are 
probably giant molecular clouds 
near the Centre. 

 From   $\approx$ 6500 input entries, cross-identifications led
to 5004 clouds and complexes in the unified catalogue, which will be available in electronic
form at CDS (Strasbourg) via anonymous ftp to {\it cdsarc.u-strasbg.fr (130.79.128.5)}.   Each entry occupies 2 catalogue lines. 
The first line shows by columns: (1) and (2) Galactic coordinates, (3) and (4) J2000.0 $\alpha,\delta$,
(5) and (6) major and minor axes dimensions in arc minutes, (7) optical opacity class, 
(8) designation(s). In line 2 columns (1) to (5)  list FIR central reddening, background
reddening, fluctuation and contrast parameters $\Delta$ and $\rho$.
Subsequent fields are distances in kpc preceded by `d=', and
CO or H\,I LSR velocities preceded by `vCO=' or `vHI=', respectively. Finally, comments occupy
the last part of line 2: (i)   further designations (preceded by '\&') with respect to 
column (8) of line 1; (ii) hyerarchical relations among dark and/or bright nebulae (e.g. LBN for  Lynds  \cite{lyn2} and Sh2- for Sharpless \cite{sha}), as  indicated by {\it in} (within), {\it inc} (includes) or  {\it rel} (related). Note that  dust cloud designations 
are not mixed to  those of bright clouds, since such possible  relations are given only in line 2.
Concerning IREC clouds we include the remarks by D\'esert et al. (\cite{des}) on those 
disturbed by bad IR or HI pixels.
Dust cloud acronyms in the catalogue other than those indicated in the present references, 
are taken from  the references themselves.

\section{Some catalogue properties and prospective work}

     Fig. 1 shows  $\it l$ and $\it b$  histograms for the resulting
5004 dust clouds. The $\it l$ histogram 
presents a minimum towards the Anticentre and  maximum in the Galactic 
Centre direction. In the  $\it b$ histogram  the clouds are strongly concentrated to the Plane, 89 \%  located in the  low-latitude zone. In the high-latitude zone  
there occur 525 clouds. A few clouds populate the Polar Caps. The present  sample of high-latitude clouds  
is  a factor $\approx$ 2 larger than that recently compiled by Bhatt (2000) 
in a  study  of  possible connection of part of them to the 
Per\,OB3/CasTau and Sco\,OB2 associations.

\begin{figure}
\centering
\resizebox{\hsize}{!}
{\includegraphics{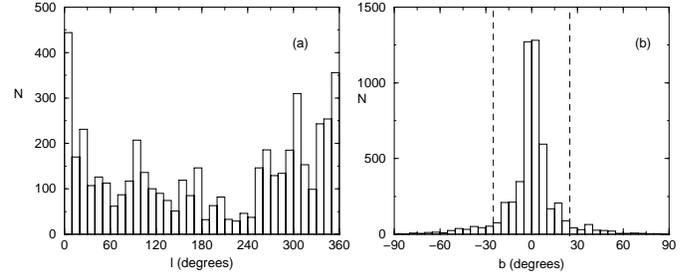}}
\caption[]{Histograms of dust clouds for Galactic longitudes 
(Panel (a)) and latitudes (Panel (b)). Dashed line separates  
adopted high and low latitude zones.} 
\label{FigGame}
\end{figure}

\begin{figure}
\centering
\resizebox{\hsize}{!}
{\includegraphics{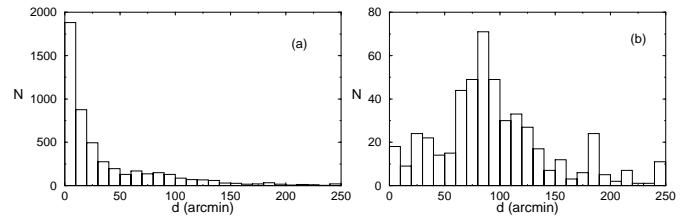}}
\caption[]{Angular size histograms for low (left panel) and high  (right  panel) 
latitude dust clouds. Major axis dimension  is used. 
The plots are limited to 250'.} 
\label{FigGame}
\end{figure}

Angular size histograms are shown in Fig. 2 in terms of cloud major axis dimension.
Each histogram is limited to 250$^{\prime}$, but some clouds and complexes
exceed 1000$^{\prime}$. For the total sample (Panel a) the distribution is exponential, 
dominated by  globules and  cloud cores.  For the high-latitude sample
(Panel b) the distribution is $\approx$ gaussian, and the peak occurs for clouds of size 80'-90'.

The present effort to merge and cross-identify dust cloud catalogues 
allows one to conclude that $\approx$ 1500 have a  previously reported 
equivalent object. Chronology (Tables 1
and 2) and usage help decide which acronym(s) to adopt. 
Just to mention a few examples: (i) Barnard's early catalogue has some
clouds overlapping with those of large southern catalogues, e.g. B 232, FeSt1-252 and HMSTG 343.7+4.0
are equivalent objects; (ii) the northern/equatorial and southern largest catalogues
have an overlapping zone, e.g the small cloud LDN 1701 = HMSTG 354.0+3.5 = Sc16,
the latter designation  is from Vilas-Boas et al. (\cite{vil2}); (iii) the small
dust cloud in the Gum Nebula GDC 6 has  counterparts in HMSTG 267.6-6.4 and BHR 40,
but its neighbour GDC 3 remains to date with a single designation; (iv) at high
latitudes MBM 2 = IREC 155 originate from 2 different methods (CO and IR/H\,I excess);
(v) IREC and DIR clouds have many equivalences, e.g. IREC 7 = DIR 009-30; (vi) catalogues
typical of high and low latitude zones have clouds in common, e.g. LDN 317 = IREC 10, and
FeSt1-223 = IREC 445.
The catalogue is not intended to be complete, but it includes 
all cross-identified acronyms, and provides 
a tool to converge information. Doubt on any cross-identification or
parameter can be analysed by accessing  the original catalogues.

Finally, we address  prospective work. A promising approach 
is the identification of dust
clouds in the $K_s$ band with 2MASS (Dutra \& Bica \cite{dut4}). The present
catalogue is a tool to verify whether NIR clouds have an optical counterpart 
or not. Recently,  2MASS  reported 
in {\it http://www.ipac.caltech.edu/2mass/gallery/fest1-457atlas.jpg} a 3-band image 
and  description of a cloud identified as FeSt1-457 
(Feitzinger \& St\"uwe \cite{fei}). This was a near-position match. A detailed 
cross-identification with all available information shows that one is dealing 
with 2 different objects. 
The 2MASS object (included as 2MASS-DN1.7+3.6 in the present catalogue) 
has a small angular size ($\approx 4^{\prime}\times 3^{\prime}$) with 
no optical counterpart on DSS or XDSS images, while FeSt1-457 is a large cloud 
($\approx 70^{\prime}$)  with  medium optical opacity (class 4).
If both clouds are nearby, that in  2MASS  may be a related globule, but it may turn out
to be a more distant object. This application shows the importance of a unified 
catalogue for searches  of new clouds.

\begin{acknowledgements}

We are indebted to Prof. Harm Habing and an anonymous referee for interesting suggestions
and remarks.
We made use of electronic catalogues from  CDS (Simbad and VizieR). We employed the Guide Star Catalog
and  Digitized Sky Survey images (DSS and XDSS), both produced at the Space Telescope Science Institute under U.S. Government grants NAS5-26555 and  NAG W-2166. The Canadian Astronomy Data Centre (CADC) interface was
used for DSS and XDSS extractions. We acknowledge support from the Brazilian institution CNPq. CD acknowledges FAPESP for a post-doc fellowship (proc. 00/11864-6).

\end{acknowledgements}

\end{document}